\documentclass[letterpaper]{article}

\usepackage[T1]{fontenc}

\usepackage{geometry}
\usepackage{setspace}

\usepackage{achemso}
\setkeys{acs}{articletitle=true, maxauthors=10}

\usepackage{amsmath}
\usepackage{amssymb}
\usepackage{bm}

\usepackage{graphicx}
\usepackage{float}
\usepackage{booktabs}
\usepackage{multirow}

\usepackage[version=4]{mhchem}
\usepackage{siunitx}

\usepackage{amsthm}

\usepackage{hyperref}
\hypersetup{colorlinks=true,linkcolor=blue,citecolor=blue,urlcolor=blue}

\usepackage{authblk}

\title{An Iterative Dual-Channel Neural Quantum State Algorithm for
  Selected Configuration Interaction}

\author[1,\dag]{Jen-Yu Chang}
\author[2,\dag]{Yi-Chun Chang}
\author[3,\dag]{Yu-Jui Lin}
\author[1,\dag]{Ming-Chun Yang}
\author[2]{Hsiu-Chi Tsai}
\author[4]{Tai-Yue Li}
\author[4]{Nan Yow Chen}
\author[5,6]{Tsung-Wei Huang}
\author[1,*]{En-Jui Kuo}

\affil[1]{Department of Electrophysics, National Yang Ming Chiao Tung University, Hsinchu, Taiwan}
\affil[2]{Arete Honors Program, National Yang Ming Chiao Tung University, Hsinchu, Taiwan}
\affil[3]{Master Program in Quantum Science and Technology, National Yang Ming Chiao Tung University, Hsinchu, Taiwan}
\affil[4]{National Center for High-performance Computing, National Institutes of Applied Research, Hsinchu, Taiwan}
\affil[5]{Quantum Information Center, Chung Yuan Christian University, Taoyuan, Taiwan}
\affil[6]{Master Program in Intelligent Computing and Big Data, Chung Yuan Christian University, Taoyuan, Taiwan}

\affil[$\dag$]{These authors contributed equally.}
\affil[*]{Corresponding author.}

\date{E-mail: \href{mailto:kuoenjui@nycu.edu.tw}{kuoenjui@nycu.edu.tw}}

\begin{document}

\maketitle

\begin{abstract}
Accurately solving the electronic Schr\"{o}dinger equation for strongly
correlated systems remains a central challenge in quantum chemistry,
where the exponential growth of configuration space limits the
applicability of exact methods.
Selected Configuration Interaction (SCI) algorithms address this
challenge by adaptively constructing compact determinantal expansions,
yet their efficiency depends critically on the quality of the sampling
strategy used to identify chemically important configurations.
Here we introduce the Handover Iterative Neural Quantum State (HI-NQS)
algorithm, which embeds a classically trained autoregressive Transformer
neural quantum state within the iterative sample--diagonalize--update
framework of Sample-Based Quantum Diagonalization.
A dual-channel Transformer architecture with explicit spin-up/spin-down
cross-attention encodes fermionic spin structure as an architectural
inductive bias, enabling expressive and physically informed wavefunction
representations.
After each subspace diagonalization, the resulting eigenvector is
distilled back into the network through a factorized spin-marginal
teacher signal, establishing a closed feedback loop between generative
sampling and exact diagonalization.
Benchmarks across a range of small molecules and a systematic
nitrogen active-space series demonstrate that HI-NQS achieves chemical
accuracy on all systems tested, with determinant-count scaling substantially
more favorable than conventional CIPSI-based SCI for all but the smallest active spaces.
All calculations are performed on GPU hardware without
quantum computing resources, establishing HI-NQS as an efficient and
scalable purely classical approach to the selected configuration
interaction problem.
\end{abstract}

  \section{\label{sec:intro}Introduction}

    Accurately computing the ground-state energy of strongly correlated
    molecular systems is a central challenge of \textit{ab initio}
    quantum chemistry, because the many-body Hilbert space grows
    exponentially with system size.
    Full Configuration Interaction (FCI) expands the wavefunction on the
    complete Slater-determinant basis~\cite{Szabo1982} and is therefore
    intractable beyond small active spaces, a bottleneck that has driven
    a succession of reduction strategies on both classical and quantum
    hardware.

    On the classical side, SCI
    algorithms~\cite{Huron1973,Holmes2016,Sharma2017,Tubman2020,Schriber2016}
    retain only the most energetically important determinants, iteratively
    growing the basis through excitation-based candidate generation and
    perturbative screening.
    Complementary approaches include the density matrix renormalization
    group (DMRG)~\cite{Chan2011}, which reaches high accuracy for
    quasi-one-dimensional correlation topologies, and FCI Quantum
    Monte Carlo (FCIQMC)~\cite{Booth2009}, which samples the complete
    determinant space stochastically without explicit enumeration.
    Modern SCI variants reach chemical accuracy on small-to-medium active
    spaces, but their candidate pool grows combinatorially with
    excitation rank, and the subspace diagonalization that follows
    scales cubically in basis size---limiting their reach into strongly
    correlated regimes where the relevant basis must grow rapidly.

    Quantum hardware has motivated a parallel lineage.
    The Variational Quantum Eigensolver (VQE)~\cite{Peruzzo2014,McClean2016} and its
    adaptive variants such as ADAPT-VQE~\cite{Grimsley2019} parameterize
    the wavefunction through a circuit, but barren
    plateaus~\cite{McClean2018,Larocca2025}, measurement overheads that
    already reach $10^{9}$--$10^{10}$ shots for small molecules at chemical
    accuracy~\cite{Wecker2015,Gonthier2022}, and irreducible gate noise
    render it impractical at scale.
    Sample-Based Quantum Diagonalization
    (SQD)~\cite{RobledoMoreno2025,Yu2025} and the related
    quantum-selected configuration interaction (QSCI)
    scheme~\cite{Kanno2024} addressed these costs with a
    \emph{hybrid} strategy: the quantum device samples bitstrings
    from the circuit distribution, while a classical computer
    constructs the second-quantized Hamiltonian restricted to
    that sampled subspace and diagonalizes it exactly.
    This separation reduces the measurement burden substantially,
    but each invocation of the quantum sampler remains a one-shot
    procedure with no mechanism to refine the sampled subspace
    across calls.

    Handover Iterative VQE (HI-VQE)~\cite{PellowJarman2025,Yoo2026} formalized and
    extended this hybrid strategy into the \emph{handover
    principle}---the defining idea captured in its name: the
    quantum device is responsible solely for proposing
    configurations, and all energy evaluation is permanently
    handed over to classical exact diagonalization.
    Crucially, HI-VQE closes a feedback loop: after each
    subspace diagonalization, the resulting eigenvector is fed
    back to update the variational circuit parameters,
    concentrating subsequent sampling on the most energetically
    relevant configurations.
    This iterative \emph{sample--diagonalize--update} cycle
    converges to chemical accuracy on systems such as N$_{2}$
    and Fe--S clusters with shot counts substantially below those of
    standard VQE.

    A parallel line attempts to realize the same selection-plus-feedback
    idea using classical Neural Quantum States (NQS)~\cite{Carleo2017,Pfau2020,Hermann2020,Choo2020,Lange2024},
    which encode the wavefunction implicitly in network parameters.
    Pure NQS variational Monte Carlo (VMC), however, relies on
    Stochastic Reconfiguration~\cite{Sorella1998}, whose Fisher-matrix inversion scales as
    $\mathcal{O}(P^{3})$ for $P$ parameters~\cite{Li2023,Chen2024}, and
    ---more fundamentally---provides no external accuracy anchor:
    whatever energy the network converges to, there is no mechanism to
    diagnose or correct its residual variational bias.
    Two complementary families of neural-network-augmented SCI close this
    gap by adding an external scoring or diagonalization step.
    The first is \emph{discriminative}: Neural-Network Configuration
    Interaction (NNCI)~\cite{Schmerwitz2025,Bilous2025,Thirion2025}, extending
    earlier neural-network-driven CI selection~\cite{Coe2018,Herzog2023},
    trains a convolutional classifier via active learning to score
    determinants drawn from an externally enumerated single- and
    double-excitation pool, reproducing the N$_{2}$ correlation energy
    with $\sim\!4\!\times\!10^{5}$ determinants in a 52-orbital active
    space where the underlying FCI dimension exceeds $10^{10}$.
    Being a classifier over an externally enumerated pool, NNCI's
    reachable configuration space is tied to the excitation rank used to
    build that pool, and its 1D-CNN architecture treats the occupation
    vector as a single $\alpha$-then-$\beta$ concatenated sequence with
    no cross-channel attention between the spin-up and spin-down blocks.

    The second family is \emph{generative}: an NQS is trained to sample
    candidate determinants directly from a learned Born distribution,
    with the CI subspace diagonalization providing the external accuracy
    anchor.
    Five recent methods fall in this family:
    \begin{itemize}
      \item \textbf{QiankunNet-cuSCI}~\cite{SuncuSCI2026} (built on the
        QiankunNet Transformer NQS framework~\cite{Wu2023NNQS,ShangQKN2025}):
        GPU-accelerated NQS+SCI scaling to active spaces of up to
        \mbox{84 qubits} (Cr$_2$); the Transformer is trained by
        variational energy minimization and candidates are proposed
        through GPU-kernel-accelerated Hamiltonian excitation enumeration.
      \item \textbf{NQS-SC}~\cite{Solanki2026}: neural-backflow ansatz on
        a selected-configuration subspace; energy minimization on the
        selected space, with new candidates drawn from the
        $\hat{H}$-connected expansion $\hat{H}\,\mathcal{S}_{\mathrm{select}}$.
      \item \textbf{HAAR-SCI}~\cite{HAAR2025}: gated Transformer that
        samples determinants autoregressively with Gumbel Top-$K$~\cite{Kool2019}
        exploration; trained by regression to Lanczos eigenvector
        amplitudes, with GPU min-heap kernels retaining only
        configurations with the largest Hamiltonian couplings.
      \item \textbf{ARNN-SCI}~\cite{Thompson2026}: masked-autoregressive
        dense (MADE-style~\cite{Germain2015}) network trained by maximum likelihood on
        configurations sampled from each iteration's SCI eigenvector;
        new configurations are then proposed by direct autoregressive
        sampling.
      \item \textbf{GTNN-SCI}~\cite{ShangGenCI2025} (also built on
        QiankunNet~\cite{Wu2023NNQS,ShangQKN2025}): generative Transformer
        regressed to CI eigenvector amplitudes; operates in two phases,
        transitioning from Hamiltonian-guided candidate scoring to direct
        autoregressive sampling once the learned distribution is
        sufficiently sharp.
    \end{itemize}
    These methods confirm that generative NQS sampling and CI subspace
    diagonalization are synergistic; three of them
    (HAAR-SCI~\cite{HAAR2025}, ARNN-SCI~\cite{Thompson2026},
    GTNN-SCI~\cite{ShangGenCI2025}) already feed CI eigenvector
    information back into the NQS through supervised losses.
    Their shared architectural limitation, however, is that
    all of these networks---whether a backflow ansatz (NQS-SC), a
    single-stream Transformer (QiankunNet-cuSCI, HAAR-SCI, GTNN-SCI), or
    a masked-dense ARNN (ARNN-SCI)---process the $\alpha$ and $\beta$
    spin-orbital indices as positions in a single sequence, with no
    dedicated cross-channel attention between the spin-up and spin-down
    subsystems.

    We propose \textbf{HI-NQS},
    which adopts the handover loop of HI-VQE~\cite{PellowJarman2025,Yoo2026}
    as its architectural backbone, substitutes the quantum-circuit
    sampler with a classically trained autoregressive Transformer NQS,
    and retains SQD as the classical subspace solver.
    Building on the second-quantized autoregressive
    ansatz~\cite{Barrett2022,Sharir2020}---which factorizes the Born distribution
    $\pi_\theta(\mathbf{x}) = \prod_i p_\theta(x_i\mid x_{<i})$ to enable
    exact ancestral sampling under hard particle-number constraints---we
    choose a Transformer rather than a recurrent or backflow backbone
    for two practical reasons: its conditionals are evaluable in parallel
    in a single forward pass, and its self-attention mechanism extends
    naturally to the cross-channel coupling described next.
    The first architectural element is a \emph{dual-channel}
    design in which a spin-$\alpha$ stream of causal self-attention
    layers is coupled to a spin-$\beta$ stream via cross-attention to
    the full $\alpha$ configuration, making the $\alpha$/$\beta$
    separation of second-quantized fermions an explicit architectural
    prior rather than one the network must learn from data; this
    contrasts directly with the single-stream encodings of all five
    generative baselines and with NNCI's concatenated 1D-CNN encoding.

    The second contribution lies in how the network is trained and how
    it explores configuration space.
    After each SQD diagonalization, the resulting eigenvector
    $\Psi_0$ is distilled into the NQS through a teacher signal given by
    the factorized product of its spin-resolved marginals,
    $p_\alpha(\sigma^\alpha)\,p_\beta(\sigma^\beta)$, optimized with
    exactly backpropagated gradients rather than by energy minimization
    alone.
    Three of the generative baselines above
    (HAAR-SCI, ARNN-SCI, GTNN-SCI) already employ supervised losses
    against CI amplitudes; the contribution here is the
    \emph{factorization} of that target into per-spin-channel marginals,
    which is architecturally consistent with the two-channel network
    of the preceding paragraph and turns the dual-channel
    cross-attention into an actively supervised pathway rather than a
    passive structural prior.
    A deterministic classical expansion of single and double excitations
    of the dominant determinants in the current basis is added in
    parallel with the NQS proposals to counteract the mode collapse~\cite{Malyshev2024}
    that afflicts autoregressive samplers at late iterations and to
    guarantee a steady supply of fresh determinants in the
    $\hat{H}$-connected vicinity of the high-amplitude region.
    Benchmarks on an eight-point N$_{2}$ active-space series sweeping
    CAS(6,6) through CAS(14,20) (12 to 40 qubits, cc-pVDZ) together with
    ten small molecules at \mbox{12--30 qubits} demonstrate that HI-NQS
    attains chemical accuracy ($|E - E_{\mathrm{ref}}| < 1.6$~mHa) on all
    eighteen systems, with the N$_{2}$ variational basis growing from
    $\sim\!4\!\times\!10^{4}$ determinants at CAS(14,14) to
    $\sim\!1.7\!\times\!10^{5}$ at CAS(14,20) (40 qubits)---roughly
    $200\times$ ($\approx 2.3$ orders of magnitude) fewer than CIPSI-SCI
    at the largest active space, and yielding a log-linear scaling
    exponent $\alpha_{\mathrm{NQS}} = 0.089 \pm 0.011$ in $N_{\mathrm{det}}$ versus qubit
    count, roughly half that of CIPSI-SCI ($\alpha_{\mathrm{CIPSI}} = 0.180$) and a third
    that of complete active-space CI (CASCI; $\alpha_{\mathrm{CASCI}} = 0.262$).
    At 40 qubits HI-NQS requires $\sim\!200\times$ fewer determinants than
    CIPSI-SCI while matching or exceeding its accuracy, all on commodity GPU
    hardware without quantum resources.

  \section{\label{sec:method}Method}

    The HI-NQS algorithm iterates between six stages:
    (i) an autoregressive Transformer NQS that
    proposes candidate Slater determinants;
    (i$'$) a deterministic classical expansion that enumerates single and
    double excitations of the dominant determinants in the current basis,
    counteracting mode collapse of the autoregressive sampler;
    (ii) a perturbation-theory scoring stage that selects the most
    energetically relevant configurations from the combined candidate
    pool;
    (iii) an SQD diagonalization that yields an
    updated ground-state energy and eigenvector;
    (iv) a supervised update in which the SQD eigenvector is distilled
    back into the NQS as a teacher signal; and
    (v) a convergence check that terminates the loop once the variational
    energy has stabilized.
    After variational convergence, the converged energy $E_{\mathrm{var}}$ is a
    strict upper bound to the exact energy and is the quantity reported throughout
    this work; no post-processing PT2 correction is applied, because the
    Epstein--Nesbet~\cite{Epstein1926,Nesbet1955} correction is non-variational---it can push the total energy
    below the exact FCI ground state at insufficient basis sizes---and CIPSI-SCI
    likewise reports bare variational energies, placing both methods on equal footing.
    Each component is described in the following subsections.

  \subsection{\label{sec:nqs_arch}Neural quantum state architecture}

  \begin{figure}[!t]
    \centering
    \includegraphics[width=\linewidth]{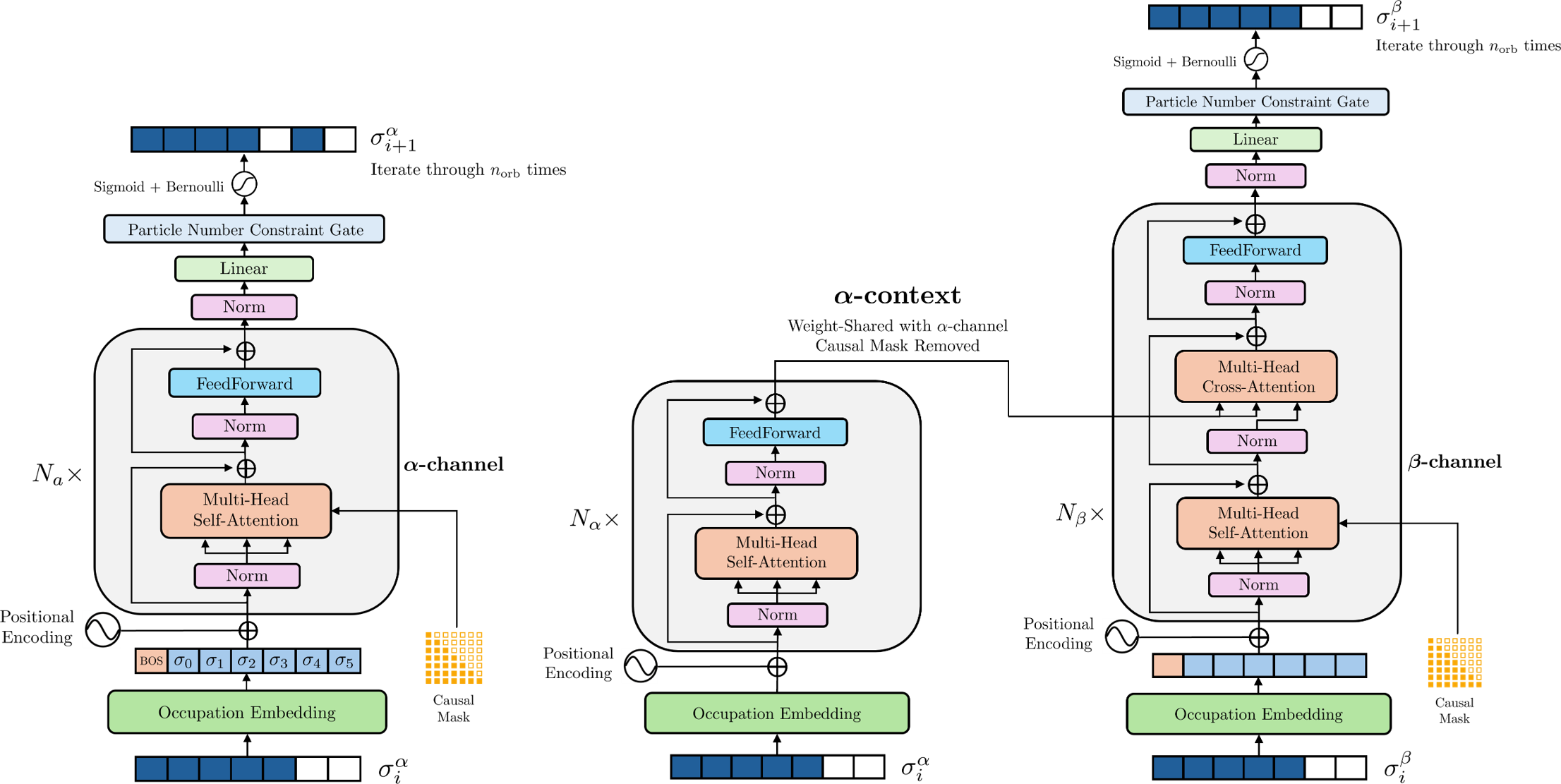}
    \caption{Dual-channel autoregressive Transformer NQS architecture.
    The spin-$\alpha$ stream (left) applies causal self-attention to generate
    spin-up conditionals; its output is re-encoded without masking to form the
    $\alpha$-context $\mathbf{H}^\alpha$.
    The spin-$\beta$ stream (right) interleaves causal self-attention with
    cross-attention to $\mathbf{H}^\alpha$, so each $\beta$ conditional has
    access to the complete spin-up configuration.
    Each channel terminates in a per-position sigmoid output head; all
    conditionals are evaluable in a single forward pass.}
    \label{fig:architecture}
  \end{figure}

    We represent the electronic wavefunction with an autoregressive
    Transformer~\cite{Vaswani2017}---related self-attention ans\"atze in
    first-quantized molecular electronic structure include
    Psiformer~\cite{vonGlehn2023}---that models the joint probability
    distribution over occupation-number strings
    $\mathbf{x} = (x_1, x_2, \ldots, x_{2M})$, where $M$ is the number of
    spatial orbitals and each $x_i \in \{0,1\}$ indicates whether
    spin-orbital $i$ is occupied.
    We organize the occupation string into a spin-$\alpha$ sub-string
    $\sigma^{\alpha}=(x_1^\alpha,\ldots,x_M^\alpha)$ and a spin-$\beta$
    sub-string $\sigma^{\beta}=(x_1^\beta,\ldots,x_M^\beta)$, so that
    $\mathbf{x}=(\sigma^{\alpha},\sigma^{\beta})$, and factorize
    the Born distribution as
    \begin{equation}
      \pi_\theta(\mathbf{x})
      = \prod_{i=1}^{M} p_\theta(x_i^\alpha \mid x_{<i}^\alpha)
      \; \prod_{j=1}^{M} p_\theta(x_j^\beta \mid x_{<j}^\beta,\,
                                  \sigma^{\alpha}),
      \label{eq:autoregressive}
    \end{equation}
    an exact chain-rule factorization in which the $\beta$ conditionals
    depend on the full $\alpha$ sub-string.
    This autoregressive form is chosen so that
    (i) $\pi_\theta(\mathbf{x})$ is normalized by construction and
    $\log\pi_\theta$ admits an analytic, partition-function-free
    evaluation;
    (ii) configurations can be drawn as independent samples by sequential
    conditional sampling, without Markov-chain mixing or rejection; and
    (iii) discrete symmetries such as particle-number conservation can be
    imposed exactly at the level of each conditional, as described below.
    Together these three properties---exact log-probabilities, unbiased
    samples, and built-in symmetry---are precisely what the perturbative
    selection step of Sec.~\ref{sec:scoring} and the supervised update of
    Sec.~\ref{sec:nqs_update} require, and they motivate the
    autoregressive architecture detailed below in preference to
    Markov-chain-based neural ans\"atze.

    The two factors in Eq.~\eqref{eq:autoregressive} are parameterized by
    two dedicated Transformer streams, which we refer to as the $\alpha$
    channel and the $\beta$ channel (Fig.~\ref{fig:architecture}).
    The $\alpha$ channel is a stack of $L_\alpha$ pre-LayerNorm~\cite{Xiong2020} Transformer
    blocks with causal (masked) multi-head self-attention that generates
    the spin-up conditionals $p_\theta(x_i^\alpha \mid x_{<i}^\alpha)$.
    The full sampled $\sigma^{\alpha}$ is then re-encoded through the
    same $\alpha$ weights without the causal mask to produce a
    bidirectional contextual representation
    $\mathbf{H}^\alpha\in\mathbb{R}^{M\times d_{\mathrm{model}}}$, which we
    refer to as the $\alpha$-context.
    The $\beta$ channel is a second stack of $L_\beta$ Transformer blocks,
    each of which augments causal self-attention over previously sampled
    $\beta$ positions with \emph{cross-attention} to the
    $\alpha$-context $\mathbf{H}^\alpha$, so that each $\beta$ conditional
    has access to the complete $\alpha$ configuration.
    Each output head produces a single logit per position, passed through
    a sigmoid to yield $p_\theta(x_i=1\mid\cdot)$, so that
    $\log\pi_\theta(\mathbf{x})$ can be evaluated in parallel over all
    positions in a single forward pass via binary cross-entropy with
    logits.

    Particle-number symmetry is enforced \emph{during} autoregressive
    sampling rather than by post-hoc rejection.
    At each sampling step $i$ the logit $\ell_i$ is clipped to a large
    positive value whenever the remaining sites in the current spin
    channel must all be occupied to reach the target electron count, and
    to a large negative value whenever the target has already been
    saturated.
    Every drawn configuration therefore satisfies
    $\sum_i x_i^{\alpha} = n_\alpha^{e}$ and
    $\sum_i x_i^{\beta} = n_\beta^{e}$ by construction; a residual
    particle-number filter is retained only as a numerical safety net.
    Fixing $(n_\alpha^{e}, n_\beta^{e})$ separately also fixes the total
    spin projection $S_z = (n_\alpha^{e} - n_\beta^{e})/2$;
    $\hat{S}^2$ is not constrained at the ansatz level and is resolved by
    the subsequent subspace diagonalization.

  \subsection{\label{sec:loop}Iterative handover loop}

  \begin{figure}[!t]
    \centering
    \includegraphics[width=\columnwidth]{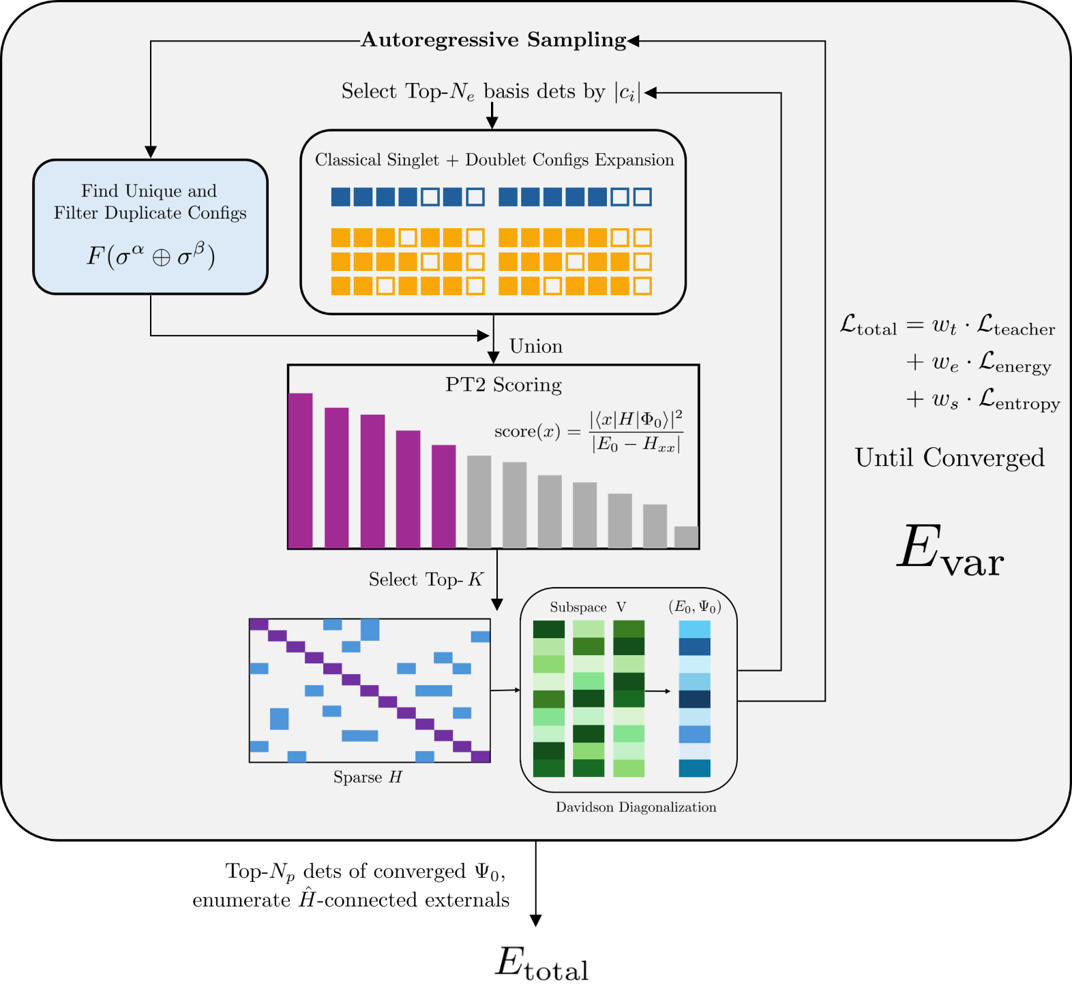}
    \caption{The iterative HI-NQS handover loop.}
    \label{fig:loop}
  \end{figure}

    Given a second-quantized molecular Hamiltonian
    $\hat{H} = \sum_{pq} h_{pq}\, \hat{a}_p^\dagger \hat{a}_q
             + \tfrac{1}{2}\sum_{pqrs} g_{pqrs}\,
               \hat{a}_p^\dagger \hat{a}_q^\dagger \hat{a}_s \hat{a}_r
             + E_{\mathrm{nuc}}$
    in an orthonormal orbital basis, the algorithm proceeds as follows
    (Fig.~\ref{fig:loop}).

    \subsubsection{Step 1: NQS sampling}
    At iteration $t$ the NQS draws $N_s$ occupation strings
    autoregressively [Eq.~\eqref{eq:autoregressive}] at a temperature
    $T(t)$ that is annealed linearly from $T_0 = 1.0$ to
    $T_{\min} = 0.3$ over the course of the run.
    The schedule biases early iterations toward broader exploration and
    later iterations toward the peak of the learned Born distribution.
    Configurations duplicating entries of the cumulative basis
    $\mathcal{B}$ are removed, yielding a set of NQS-proposed
    candidates $\mathcal{C}_t^{\mathrm{NQS}}$.

    \subsubsection{Step 1$'$: Classical expansion}
    \label{sec:classical_expansion}
    In addition to the stochastic NQS proposals, a deterministic
    expansion is performed every iteration once an eigenvector is
    available ($t \geq 1$).
    Let $\mathcal{B}_{\text{top}}^{(N_e)} \subset \mathcal{B}$ denote the
    $N_e$ basis determinants with the largest amplitudes
    $|c_i|$ in the current eigenvector $\Psi_0$.
    We enumerate every singly and doubly excited determinant connected
    to any element of $\mathcal{B}_{\text{top}}^{(N_e)}$ by a non-zero
    Hamiltonian matrix element,
    \begin{equation}
      \mathcal{C}_t^{\mathrm{cl}}
      = \big\{\, \mathbf{y} \notin \mathcal{B} \,:\,
        \exists\, \mathbf{x} \in \mathcal{B}_{\text{top}}^{(N_e)},\;
        \langle \mathbf{y}|\hat{H}|\mathbf{x}\rangle \neq 0 \,\big\},
      \label{eq:classical_expansion}
    \end{equation}
    and merge $\mathcal{C}_t^{\mathrm{cl}}$ with $\mathcal{C}_t^{\mathrm{NQS}}$
    to form the full candidate pool
    $\mathcal{C}_t = \mathcal{C}_t^{\mathrm{NQS}} \cup \mathcal{C}_t^{\mathrm{cl}}$.
    This step compensates for the gradual collapse of $\pi_\theta$ onto
    high-amplitude regions of $\mathcal{B}$ at later iterations: even when
    the NQS proposal distribution becomes peaked, the classical expansion
    guarantees a steady supply of fresh determinants in the
    $\hat{H}$-connected vicinity of the dominant amplitudes.
    We use $N_e = 1000$ throughout this work.

    \subsubsection{Step 2: Configuration scoring and basis update}
    \label{sec:scoring}
    Candidates are ranked by their expected contribution to the
    ground-state energy, and the cumulative basis $\mathcal{B}$ is updated
    accordingly.
    The ranking criterion depends on whether an approximate eigenvector
    is available from a previous diagonalization.

    \paragraph{Iteration $t=0$ (cold start).}
    No eigenvector $\Psi_0$ is yet available.
    Candidates are ranked by their diagonal Hamiltonian matrix element
    $H_{\mathbf{x}\mathbf{x}} = \langle\mathbf{x}|\hat{H}|\mathbf{x}\rangle$.
    The $K_0$ lowest-energy candidates form the initial basis
    $\mathcal{B}$, with the cold-start budget $K_0$ chosen no larger than
    the per-iteration budget $K$ used in later iterations.

    \paragraph{Iteration $t=1$ (full rescore).}
    Once the first eigenvector $\Psi_0$ is available, both existing
    entries in $\mathcal{B}$ and new candidates $\mathcal{C}_1$ are
    rescored using the Epstein--Nesbet second-order perturbation theory
    (PT2) importance measure
    \begin{equation}
      s(\mathbf{x})
      = \frac{|\langle \mathbf{x} | \hat{H} | \Psi_0 \rangle|^2}
             {|E_0 - H_{\mathbf{x}\mathbf{x}}|},
      \label{eq:pt2_score}
    \end{equation}
    where the coupling
    $\langle\mathbf{x}|\hat{H}|\Psi_0\rangle
     = \sum_i c_i\, H_{\mathbf{x},\mathbf{x}_i}$
    is evaluated over the eigenvector coefficients $\{c_i\}$ and basis
    determinants $\{\mathbf{x}_i\}$.
    The absolute value in the denominator makes
    $s(\mathbf{x}) \geq 0$ a magnitude-only ranking criterion.
    The basis is then rebuilt from scratch: $\mathcal{B}$ is replaced by the
    top $K$ configurations ranked by $s(\mathbf{x})$.
    The hard cap $\mathcal{B}_{\max}$ is not applied at this step so that every
    cold-start determinant---which was selected by a weaker diagonal
    criterion---receives at least one PT2 evaluation before being
    eligible for eviction.

    \paragraph{Iterations $t \geq 2$ (incremental update).}
    Only new candidates $\mathcal{C}_t$ are scored via
    Eq.~\eqref{eq:pt2_score}; the top $K$ are appended to $\mathcal{B}$.
    If $|\mathcal{B}| > \mathcal{B}_{\max}$, the lowest-scored entries are evicted.

    \subsubsection{Step 3: Subspace diagonalization}
    The projected Hamiltonian $H_{\mathcal{B}}$ with matrix elements
    $[H_{\mathcal{B}}]_{ij} = \langle\mathbf{x}_i|\hat{H}|\mathbf{x}_j\rangle$
    is constructed and diagonalized with a Davidson eigensolver~\cite{Davidson1975}.
    By default we use an incremental SQD backend that persists a PySCF~\cite{Sun2020PySCF}
    \texttt{selected\_ci} state across iterations and warm-starts
    Davidson from the previous eigenvector, with IBM's
    \texttt{solve\_fermion} routine~\cite{RobledoMoreno2025} retained as
    a fallback.
    This yields the total ground-state energy
    $E_0 = E_{\mathrm{corr}} + E_{\mathrm{nuc}}$ in the selected subspace
    and the corresponding eigenvector $\Psi_0$ expressed
    as a linear combination of the basis determinants.
    Because this step is an exact diagonalization within $\mathcal{B}$,
    $E_0$ is variational with respect to the selected subspace.

    \subsubsection{Step 4: NQS update}
    \label{sec:nqs_update}
    The NQS is retrained to reflect the eigenvector $\Psi_0$ obtained in
    Step~3.
    Training uses a composite loss
    \begin{equation}
      \mathcal{L}_{\mathrm{total}}
      = w_{\mathrm{teach}} \mathcal{L}_{\mathrm{teacher}}
      + w_{\mathrm{ener}} \mathcal{L}_{\mathrm{energy}}
      + w_{\mathrm{entr}} \mathcal{L}_{\mathrm{entropy}},
      \label{eq:loss}
    \end{equation}
    whose components are defined below.

    \paragraph{Teacher weights.}
    We compute spin-resolved marginals of $|\Psi_0|^2$,
    \begin{align}
      p_\alpha(\sigma^{\alpha}) &=
        \sum_{\sigma^{\beta}}
        |\Psi_0(\sigma^{\alpha},\sigma^{\beta})|^2,
        \label{eq:alpha_marginal}\\
      p_\beta(\sigma^{\beta}) &=
        \sum_{\sigma^{\alpha}}
        |\Psi_0(\sigma^{\alpha},\sigma^{\beta})|^2,
        \label{eq:beta_marginal}
    \end{align}
    and assign each basis configuration
    $\mathbf{x}=(\sigma^{\alpha},\sigma^{\beta})$ the factorized
    teacher weight
    $w(\mathbf{x}) = p_\alpha(\sigma^{\alpha})\,p_\beta(\sigma^{\beta})$,
    normalized so that $\sum_{\mathbf{x}\in\mathcal{B}} w(\mathbf{x}) = 1$.
    This factorized form discards the phase of $\Psi_0$ and the residual
    $\alpha$--$\beta$ correlation of $|\Psi_0|^2$.

    \paragraph{Loss components.}
    The three terms are
    \begin{align}
      \mathcal{L}_{\mathrm{teacher}}
        &= -\!\sum_{\mathbf{x} \in \mathcal{B}}
           w(\mathbf{x})\,\log\pi_\theta(\mathbf{x}),
        \label{eq:teacher} \\
      \mathcal{L}_{\mathrm{energy}}
        &= \sum_{\mathbf{x} \in \mathcal{B}}
           w(\mathbf{x})\,A(\mathbf{x})\,\log\pi_\theta(\mathbf{x}),
        \label{eq:energy} \\
      \mathcal{L}_{\mathrm{entropy}}
        &= \frac{1}{|\mathcal{B}|}\sum_{\mathbf{x} \in \mathcal{B}}
           \log\pi_\theta(\mathbf{x}),
        \label{eq:entropy}
    \end{align}
    where $A(\mathbf{x}) = H_{\mathbf{x}\mathbf{x}} - E_0$ is the
    single-determinant advantage relative to the current ground-state
    energy.
    $\mathcal{L}_{\mathrm{teacher}}$ is the cross-entropy between the
    teacher weights $w(\mathbf{x})$ and $\pi_\theta(\mathbf{x})$
    evaluated on $\mathcal{B}$.
    $\mathcal{L}_{\mathrm{energy}}$ is a REINFORCE-style policy-gradient~\cite{Williams1992}
    term that raises $\log\pi_\theta(\mathbf{x})$ for $A<0$ and lowers it
    for $A>0$.
    $\mathcal{L}_{\mathrm{entropy}}$ is an anti-concentration regularizer:
    minimizing it lowers the average log-probability on $\mathcal{B}$,
    redistributing probability mass toward configurations outside the current basis.
    Gradients are computed on mini-batches of size
    $\min(5000,|\mathcal{B}|)$ and clipped to a maximum norm of $1.0$;
    parameter updates use Adam~\cite{KingmaBa2014} with exactly backpropagated gradients.

    \subsubsection{Step 5: Convergence}
    The variational loop terminates when the energy change
    $|E_0^{(t)} - E_0^{(t-1)}|$ falls below a threshold
    $\delta = 10^{-9}\,\mathrm{Ha}$ for five consecutive iterations,
    or when a maximum number of iterations is reached.
    We use a maximum of $50$ iterations throughout this work.
    We denote the converged variational energy by
    $E_{\mathrm{var}} \equiv \lim_{t\to\infty} E_0^{(t)}$.

  \section{Results and Discussion}
  \label{sec:results}

\begin{figure}[!htbp]
      \centering
      \includegraphics[width=0.7\linewidth]{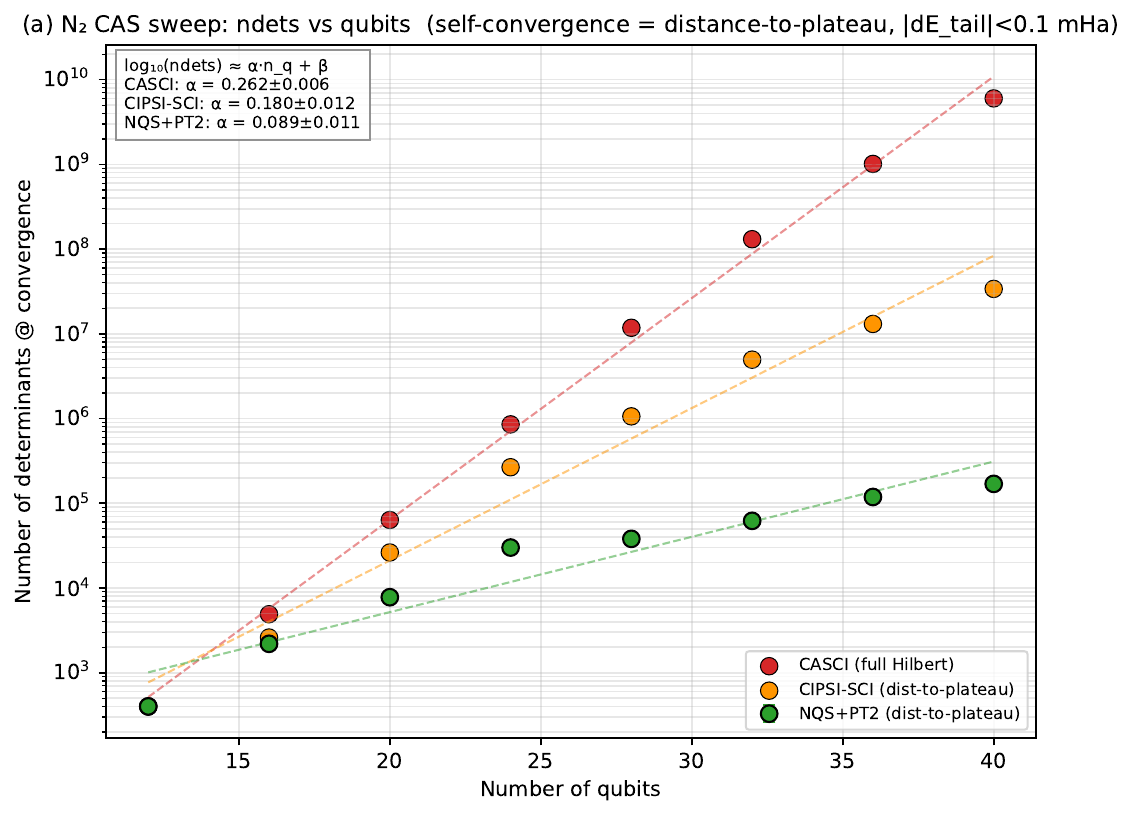}
      \caption{Determinant count at natural convergence vs.\ qubit count for the
      N$_2$ CAS series. Full CASCI dimension (red), CIPSI-SCI (orange), NQS
      variational (green); dashed lines are log-linear fits
      $\log_{10} N_\mathrm{det} = \alpha_X\, n_q + \beta_X$ (slope $\alpha_X$ per method)
      with $\alpha_\mathrm{CASCI} = 0.262 \pm 0.006$, $\alpha_\mathrm{CIPSI} = 0.180 \pm 0.012$,
      $\alpha_\mathrm{NQS} = 0.089 \pm 0.011$. Error bars on NQS span the three-seed
      min--max.}
      \label{fig:scaling}
      \end{figure}

      Across 18 benchmark systems spanning $12$--$40$ qubits, HI-NQS attains
      chemical accuracy on every system tested, with a log-linear $N_\mathrm{det}$
      scaling exponent of $\alpha_{\mathrm{NQS}} = 0.089 \pm 0.011$---half that of CIPSI-SCI
      ($\alpha_{\mathrm{CIPSI}} = 0.180$) and a third that of CASCI ($\alpha_{\mathrm{CASCI}} = 0.262$,
      Fig.~\ref{fig:scaling}). At the largest active space tested (CAS(14,20),
      40 qubits) this compounds to a $\sim\!200\times$ ($\approx 2.3$ orders
      of magnitude) determinant reduction relative to CIPSI-SCI.

\begin{figure}[!htbp]
      \centering
      \includegraphics[width=\linewidth]{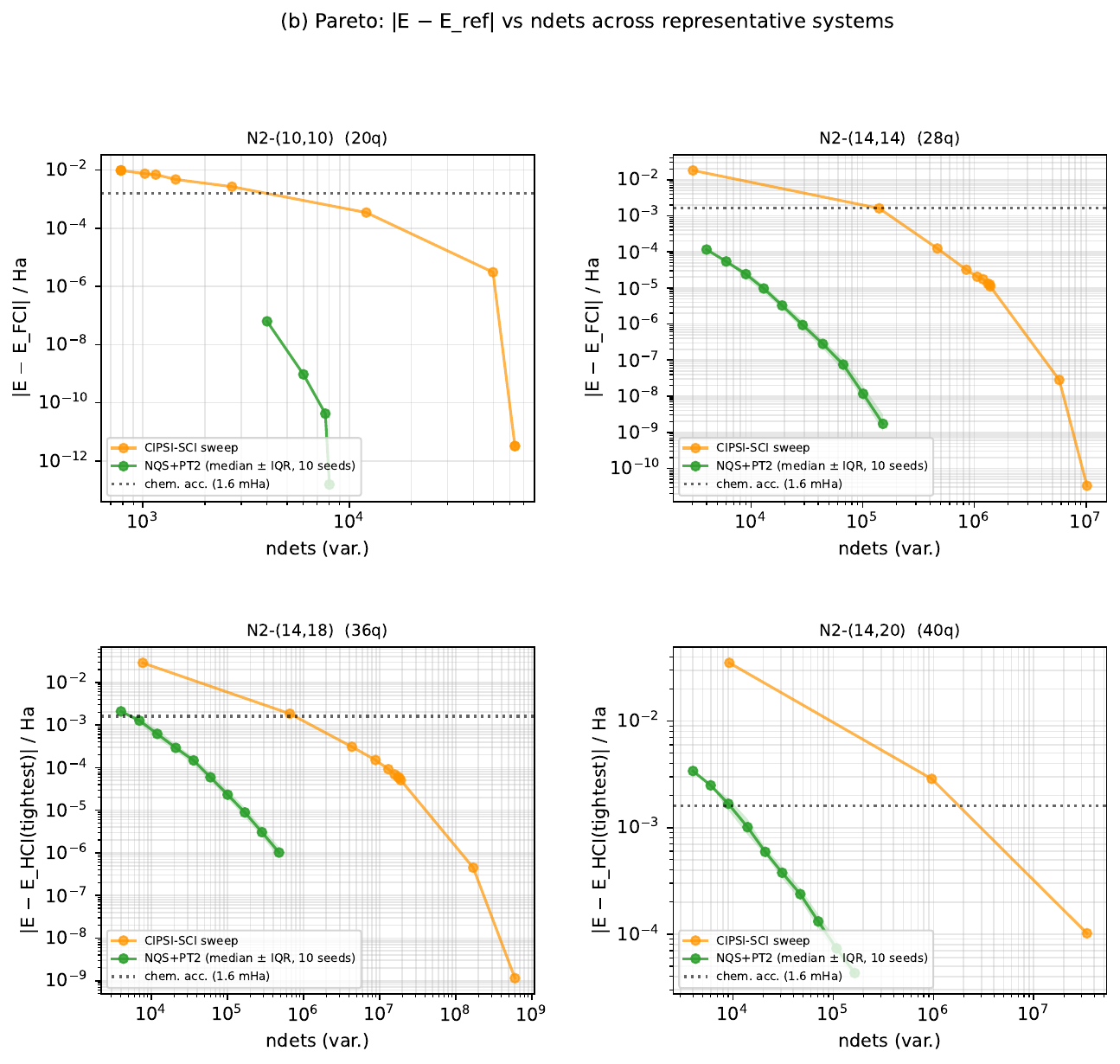}
      \caption{Accuracy--cost Pareto fronts: $|E - E_\mathrm{ref}|$ vs.\
      determinant count for four representative active spaces. CIPSI-SCI sweep
      (orange) and NQS variational (green, median $\pm$ IQR over ten seeds);
      the dotted line marks chemical accuracy ($1.6$~mHa). References are
      exact PySCF FCI except for CAS(14,20), where a Dice-SHCI+det-PT2
      estimate ($\varepsilon = 10^{-6}$) serves as an FCI proxy.}
      \label{fig:pareto}
      \end{figure}

      We benchmark HI-NQS against CIPSI-SCI (implemented via PySCF's \texttt{selected\_ci} module~\cite{Sun2020PySCF,Huron1973}) and, where tractable,
      against the exact FCI limit on two families:
      ten small molecules at $12$--$30$ qubits (Group A) and a single
      N$_2$ active-space series, CAS(6,6) through CAS(14,20), spanning $12$--$40$
      qubits in the cc-pVDZ basis (Group B). For each method we report its
      \emph{natural-convergence} determinant count $N_\mathrm{det}$: the smallest
      basis (ordered by size) from which the energy stays within $0.1$~mHa of its own
      converged plateau. This self-convergence criterion is applied identically to
      CIPSI-SCI and to the NQS variational basis, so the two are compared at
      matched convergence rather than against a fixed external target. Both methods
      report bare variational energies: CIPSI uses the subspace-diagonalization
      energy at its converged iteration, and NQS reports $E_\mathrm{var}$
      (no Epstein--Nesbet PT2 correction), a strict upper bound to the exact energy.
      Reference energies are exact FCI for
      seventeen of the eighteen systems ($E_\mathrm{ref} = E_\mathrm{FCI}$),
      computed with PySCF's Davidson FCI solver (\texttt{pyscf.fci.direct\_spin1}). The sole exception is
      N$_2$-CAS(14,20) (40 qubits, $\sim\!6\times10^9$ determinants), where
      exact FCI is computationally infeasible; we substitute a
      Dice semistochastic heat-bath CI (SHCI)+det-PT2
      reference~\cite{Holmes2016,Sharma2017} (heat-bath variational selection with
      deterministic Epstein--Nesbet PT2), $E_\mathrm{ref} = E_\mathrm{var}^{\mathrm{SHCI}} + E_\mathrm{PT2}$,
      converged across a seven-point cutoff sweep $\varepsilon \in [10^{-3}, 10^{-6}]$.
      Selected configuration-interaction at this cutoff is generally converged to
      within $\lesssim 10^{-5}$~Ha of exact FCI, two orders below the
      $1.6\times 10^{-3}$~Ha chemical-accuracy scale of any quantitative claim in
      this work. NQS energies are averaged over three
      seeds in Table~\ref{tab:table1} and Fig.~\ref{fig:scaling}, and over ten
      seeds in Fig.~\ref{fig:pareto} (median $\pm$ IQR).
      Table~\ref{tab:table1} collects the comparison.

      \begin{table}[!htbp]
      \centering
      \caption{NQS vs.\ CIPSI-SCI at natural convergence;
      the convergence criterion and reference conventions are defined in the main
      text. Columns: $E$ are converged energies (Ha), $N_\mathrm{det}$ the
      corresponding determinant counts, $N_\mathrm{NQS}/N_\mathrm{CIPSI}$ the
      compactness ratio (${<}1$: NQS more compact),
      $N_\mathrm{det}^{\mathrm{Ref}}$ the reference (FCI) Hilbert dimension, and
      $\Delta E \equiv E_\mathrm{method}-E_\mathrm{ref}$ (mHa). \textbf{Bold}
      $N_\mathrm{NQS}/N_\mathrm{CIPSI}$ and $\Delta E_\mathrm{NQS}$ mark the
      Pareto-dominant systems (NQS more compact \emph{and} at least as accurate).
      Entries shown as $0.000$ lie below the $\sim\!10^{-5}$~mHa numerical-noise
      floor (sign not meaningful). $^\star$Near-exact-FCI Dice-SHCI surrogate
      with deterministic Epstein--Nesbet PT2 ($\varepsilon=10^{-6}$; see text).}
      \label{tab:table1}
      \resizebox{\linewidth}{!}{%
      \begin{tabular}{l c l l r r r r c r r r}
      \toprule
      Molecule & Qubits & Basis & Ref. &
      $E_\mathrm{CIPSI}$ & $N_\mathrm{det}^{\mathrm{CIPSI}}$ &
      $E_\mathrm{var}$ & $N_\mathrm{det}^{\mathrm{NQS}}$ &
      $\frac{N_\mathrm{NQS}}{N_\mathrm{CIPSI}}$ & $N_\mathrm{det}^{\mathrm{Ref}}$ &
      $\Delta E_\mathrm{CIPSI}$ & $\Delta E_\mathrm{NQS}$ \\
       & & & & (Ha) & & (Ha) & & & & (mHa) & (mHa) \\
      \midrule
      \multicolumn{12}{l}{\textit{Group A --- small molecules}}\\
      LiH & 12 & STO-3G & FCI & $-7.882311$ & 35 & $-7.882323$ & 32 & \textbf{0.91} & 225 & $+0.013$ & $\mathbf{+0.001}$ \\
      BeH$_2$ & 14 & STO-3G & FCI & $-15.595118$ & 169 & $-15.595096$ & 40 & 0.24 & 1{,}225 & $0.000$ & $+0.022$ \\
      CH$_4$ & 18 & STO-3G & FCI & $-39.805955$ & 629 & $-39.805999$ & 459 & \textbf{0.73} & 15{,}876 & $+0.080$ & $\mathbf{+0.036}$ \\
      N$_2$ & 20 & STO-3G & FCI & $-107.654122$ & 1{,}088 & $-107.654097$ & 289 & 0.27 & 14{,}400 & $0.000$ & $+0.025$ \\
      HCN & 22 & STO-3G & FCI & $-91.842150$ & 2{,}333 & $-91.842152$ & 1{,}247 & \textbf{0.53} & 108{,}900 & $+0.054$ & $\mathbf{+0.052}$ \\
      H$_2$S & 22 & STO-3G & FCI & $-394.354742$ & 865 & $-394.354670$ & 57 & 0.066 & 3{,}025 & $0.000$ & $+0.071$ \\
      C$_2$H$_2$ & 24 & STO-3G & FCI & $-76.024487$ & 3{,}302 & $-76.024475$ & 1{,}740 & 0.53 & 627{,}264 & $+0.087$ & $+0.099$ \\
      H$_2$O & 26 & 6-31G & FCI & $-76.120963$ & 11{,}001 & $-76.120999$ & 6{,}006 & \textbf{0.55} & 1{,}656{,}369 & $+0.096$ & $\mathbf{+0.060}$ \\
      C$_2$H$_4$ & 28 & STO-3G & FCI & $-77.235243$ & 16{,}001 & $-77.235259$ & 9{,}009 & \textbf{0.56} & 9{,}018{,}009 & $+0.099$ & $\mathbf{+0.083}$ \\
      NH$_3$ & 30 & 6-31G & FCI & $-56.292913$ & 36{,}001 & $-56.293111$ & 15{,}015 & \textbf{0.42} & 9{,}018{,}009 & $+0.294$ & $\mathbf{+0.096}$ \\
      \midrule
      \multicolumn{12}{l}{\textit{Group B --- N$_2$ active-space scan (cc-pVDZ)}}\\
      N$_2$-CAS(6,6) & 12 & cc-pVDZ & FCI & $-109.090227$ & 400 & $-109.090227$ & 400 & 1.0 & 400 & $0.000$ & $0.000$ \\
      N$_2$-CAS(8,8) & 16 & cc-pVDZ & FCI & $-109.111247$ & 2{,}601 & $-109.111307$ & 2{,}193 & \textbf{0.84} & 4{,}900 & $+0.060$ & $\mathbf{0.000}$ \\
      N$_2$-CAS(10,10) & 20 & cc-pVDZ & FCI & $-109.125447$ & 26{,}244 & $-109.125509$ & 7{,}784 & \textbf{0.30} & 63{,}504 & $+0.062$ & $\mathbf{0.000}$ \\
      N$_2$-CAS(12,12) & 24 & cc-pVDZ & FCI & $-109.142317$ & 266{,}256 & $-109.142326$ & 30{,}000 & \textbf{0.11} & 853{,}776 & $+0.009$ & $\mathbf{0.000}$
  \\
      N$_2$-CAS(14,14) & 28 & cc-pVDZ & FCI & $-109.156677$ & 1{,}060{,}900 & $-109.156697$ & 38{,}000 & \textbf{0.036} & 11{,}778{,}624 & $+0.020$ &
  $\mathbf{0.000}$ \\
      N$_2$-CAS(14,16) & 32 & cc-pVDZ & FCI & $-109.115858$ & 4{,}955{,}076 & $-109.115880$ & 62{,}000 & \textbf{0.013} & $1.31{\times}10^{8}$ & $+0.025$ &
  $\mathbf{+0.003}$ \\
      N$_2$-CAS(14,18) & 36 & cc-pVDZ & FCI & $-109.173146$ & 13{,}075{,}456 & $-109.173221$ & 118{,}667 & \textbf{0.0091} & $1.01{\times}10^{9}$ & $+0.092$
  & $\mathbf{+0.016}$ \\
      N$_2$-CAS(14,20) & 40 & cc-pVDZ & SHCI$^\star$ & $-109.197918$ & 33{,}907{,}329 & $-109.197978$ & 169{,}333 & \textbf{0.0050} & $6.01{\times}10^{9}$ &
  $+0.102$ & $\mathbf{+0.041}$ \\
      \bottomrule
      \end{tabular}
      }
      \end{table}

      \emph{Small molecules (Group A).} All ten systems reach chemical accuracy,
      $|E - E_\mathrm{ref}| < 1.6$~mHa, with both methods, and the NQS variational
      basis is more compact than CIPSI-SCI on all ten. The gains are largest for
      the weakly correlated cases---H$_2$S ($0.066\times$, a $15\times$ reduction),
      BeH$_2$ ($0.24\times$), and N$_2$ ($0.27\times$)---and remain substantial
      for the more strongly correlated systems: HCN and C$_2$H$_2$ ($0.53\times$
      each), H$_2$O ($0.55\times$), C$_2$H$_4$ ($0.56\times$), and NH$_3$
      ($0.42\times$). The smallest advantage is LiH ($0.91\times$), where the
      dominant determinants are few and both methods converge at near-identical
      basis sizes. All ten NQS variational energies lie above the exact FCI
      ($\Delta E_\mathrm{NQS} \geq 0$), with the largest overshoot
      $+0.099$~mHa at C$_2$H$_2$---well within chemical accuracy.

      \emph{Active-space scaling (Group B).} The N$_2$ series isolates the size
      dependence. Figure~\ref{fig:scaling} plots $N_\mathrm{det}$ at convergence
      against qubit count; all three methods grow log-linearly,
      $\log_{10} N_\mathrm{det} \approx \alpha_X\, n_q + \beta_X$, with the
      NQS exponent roughly half that of CIPSI-SCI and a third that of full
      CASCI (fitted values in Fig.~\ref{fig:scaling} caption). The determinant
      advantage is therefore not a constant prefactor but \emph{compounds} with
      active-space size: from parity at CAS(6,6) (both methods saturate the
      $400$-determinant FCI space) the NQS/CIPSI ratio falls monotonically
      through $0.30$ at CAS(10,10), $0.036$ at CAS(14,14), to
      $5.0\times10^{-3}$ at CAS(14,20)---a $\sim\!200\times$ ($\approx 2.3$
      orders of magnitude) determinant reduction relative to CIPSI-SCI, and
      $\sim\!4.5$ orders below the full CASCI dimension, at $40$ qubits (for
      context, the current frontier of distributed FCI implementations reaches
      the trillion-determinant scale~\cite{Gao2024}). For CAS(14,16) and
      CAS(14,18) the reference is exact PySCF FCI on Hilbert spaces of
      $1.3\times10^8$ and $1.0\times10^9$ determinants, respectively;
      for CAS(14,20) ($6.0\times10^9$ determinants) exact FCI is infeasible
      at single-node scale and we substitute a Dice-SHCI+det-PT2 estimate at
      $\varepsilon = 10^{-6}$ as an FCI proxy; see caption of
      Table~\ref{tab:table1} for details. The
      $N_\mathrm{NQS}/N_\mathrm{CIPSI}$ ratios in Table~\ref{tab:table1}
      therefore compare two independent natural-convergence picks at every
      qubit count.

      \emph{Accuracy--cost Pareto front.} Figure~\ref{fig:pareto} shows the
      full trade-off for four representative sizes (CAS(10,10), (14,14),
      (14,18), and (14,20)). At every active space shown, the NQS curve lies
      to the left of the CIPSI-SCI sweep---it reaches any given accuracy with
      fewer determinants---and the horizontal separation widens monotonically
      from CAS(10,10) to CAS(14,20), the per-system manifestation of the
      exponent gap in Fig.~\ref{fig:scaling}. At the largest active spaces
      both curves approach a sub-mHa accuracy floor; all NQS variational
      energies sit above their respective references, consistent with the
      variational upper-bound property of $E_\mathrm{var}$.

      \emph{A monotonically size-dependent advantage.} Taken together, the
      data show that the NQS compactness gain is present across the full
      benchmark and grows monotonically with active-space size. On 13 of the
      18 systems this advantage is \emph{Pareto-dominant}: HI-NQS uses fewer
      determinants \emph{and} lands at least as close to the reference (bold
      entries in Table~\ref{tab:table1}); on the four remaining small molecules
      (BeH$_2$, N$_2$, C$_2$H$_2$, H$_2$S) it is still more compact but slightly
      less accurate, its energy trailing CIPSI by less than $0.1$~mHa and always
      within chemical accuracy.
      The sole neutral point in the entire benchmark is CAS(6,6), where both
      methods saturate the $400$-determinant FCI space. Across Group~A the advantage
      ranges from near-parity at LiH ($0.91\times$) to $15\times$ at H$_2$S,
      with NQS more compact on all ten systems. In Group~B the NQS/CIPSI
      ratio falls continuously from $0.84$ at CAS(8,8) through $0.036$ at
      CAS(14,14) to $5.0\times10^{-3}$ at CAS(14,20). The learned,
      physics-supervised NQS sampler concentrates amplitude into a compact
      support that grows far more slowly ($\alpha_{\mathrm{NQS}} = 0.089$) than CIPSI's
      perturbatively ordered expansion ($\alpha_{\mathrm{CIPSI}} = 0.180$); the compactness
      gain is thus a scaling phenomenon that becomes decisive precisely in
      the strongly correlated, large-active-space regime these methods target.

      \emph{Relation to other generative neural selected-CI.} HI-NQS shares
      the generative selected-CI strategy of several recent neural
      ans\"atze---QiankunNet-cuSCI~\cite{SuncuSCI2026},
      NQS-SC~\cite{Solanki2026}, HAAR-SCI~\cite{HAAR2025},
      ARNN-SCI~\cite{Thompson2026} and
      GTNN-SCI~\cite{ShangGenCI2025}---all of which sample determinants from
      a learned distribution and diagonalize the selected subspace. It
      differs in the two architectural respects developed in
      Sec.~\ref{sec:method}: a dual-channel $\alpha/\beta$ Transformer whose
      $\beta$ stream cross-attends to the full $\alpha$ configuration,
      encoding opposite-spin correlation as an explicit inductive bias rather
      than as a single flattened occupation string; and a supervised
      distillation loss that feeds the exact subspace eigenvector back through
      a factorized spin-marginal teacher. We attribute the shallow
      determinant-scaling exponent reported here to this combination, although
      a controlled head-to-head on identical active spaces---not recoverable
      from the published numbers of those works---would be required to isolate
      each contribution.

      \emph{Limitations.} Several caveats bound these claims.
      Table~\ref{tab:table1} reports the bare variational NQS energy
      $E_\mathrm{var}$ (no PT2 correction), which is a strict upper bound to
      the exact energy. Consequently $\Delta E_\mathrm{NQS} \geq 0$ on every
      exact-FCI reference row (seventeen of the eighteen systems), with the
      largest deviation $+0.099$~mHa at C$_2$H$_2$---within chemical
      accuracy. The sole non-exact-FCI row---N$_2$-CAS(14,20)
      (Dice-SHCI+det-PT2 proxy at $\varepsilon = 10^{-6}$)---also has
      positive $\Delta E_\mathrm{NQS}$ ($+0.041$~mHa) against the substitute
      reference; since selected configuration-interaction at this cutoff is
      generally converged to within $\lesssim 10^{-5}$~Ha of exact
      FCI~\cite{Holmes2016,Sharma2017}, the chemical-accuracy claim is
      unaffected. The CAS(14,20) row caps the verifiable accuracy of
      \emph{any} method at the proxy's residual ($\sim\!10^{-5}$~Ha), still
      two orders below our chemical-accuracy claim. The autoregressive sampler
      is prone to late-iteration mode collapse, which the deterministic
      single/double classical expansion mitigates but does not eliminate; and
      total spin $S^2$ is not constrained at the ansatz level, being restored
      only by the subspace diagonalization. Finally, all benchmarks are at a
      single molecular geometry; dissociation curves and broken-symmetry
      transition-metal systems remain to be tested.

  \clearpage
  \section{\label{sec:conclusion}Conclusion}

  We have introduced HI-NQS, a handover iterative algorithm that replaces the
  quantum-circuit sampler of HI-VQE~\cite{PellowJarman2025,Yoo2026} with a
  classically trained autoregressive Transformer and retains
  SQD~\cite{RobledoMoreno2025} as the exact classical subspace solver. The
  two defining architectural elements---a dual-channel $\alpha$/$\beta$
  cross-attention architecture that encodes opposite-spin correlation as an
  explicit inductive bias, and a supervised distillation loop that feeds
  each SQD eigenvector back into the network through a factorized
  spin-marginal teacher signal---together close the feedback loop between
  generative sampling and exact diagonalization that open-loop NQS+SCI
  methods leave open. A deterministic classical expansion of single and
  double excitations of the dominant determinants is run in parallel with
  the NQS proposals to ensure a steady supply of fresh configurations in
  the $\hat{H}$-connected neighborhood at late iterations.

  Benchmarks on ten small molecules and an eight-point \ce{N2} active-space
  series demonstrate that this design yields a systematic and
  compounding computational advantage over CIPSI-SCI~\cite{Huron1973}.
  HI-NQS achieves chemical accuracy on all ten Group~A systems and on the
  full Group~B active-space series up to CAS(14,20), while the log-linear
  scaling exponent of the required determinant count,
  $\alpha_{\mathrm{NQS}} = 0.089 \pm 0.011$, is roughly half that of CIPSI-SCI
  ($\alpha_{\mathrm{CIPSI}} = 0.180$) and a third that of CASCI ($\alpha_{\mathrm{CASCI}} = 0.262$). At 40
  qubits, HI-NQS requires roughly $200\times$ ($\approx 2.3$ orders of
  magnitude) fewer determinants than CIPSI-SCI and approximately $4.5$
  orders of magnitude fewer than the full CASCI Hilbert space, with
  HI-NQS dominating the accuracy--cost trade-off across all active spaces
  above CAS(6,6), where the two methods tie at the saturated FCI space.

  Three near-term applications emerge directly from the benchmark results. The
  favorable scaling exponent ($\alpha_{\mathrm{NQS}} = 0.089$) suggests the method can
  extend to substantially larger active spaces with manageable determinant
  growth, and the dual-channel architecture admits natural extensions to
  several settings of current interest.
  (i) \emph{Strongly correlated transition-metal chemistry.} Fe--S
  clusters~\cite{Yoo2026,RobledoMoreno2025} and metal dimers such as
  Cr$_2$~\cite{SuncuSCI2026} are canonical multireference benchmarks where
  the compactness gain should be most consequential---the determinant tail
  exhibits the strongest combinatorial growth there, precisely the regime
  in which HI-NQS's scaling advantage compounds fastest.
  The recent classical solution of the FeMo-cofactor model to chemical
  accuracy via DMRG and coupled cluster~\cite{Zhai2026} redefines the
  frontier for tractable active-space targets; the $\sim\!76$-orbital
  active space treated there is squarely within the regime where
  HI-NQS's compounding determinant advantage should be decisive.
  (ii) \emph{Larger Gaussian basis sets.} Extending from the cc-pVDZ basis
  used here to cc-pVTZ and beyond would test whether the scaling exponent
  carries through toward the complete basis-set limit.
  (iii) \emph{Bond-breaking and dissociation studies.} Potential
  energy surface scans through dissociation regimes are a natural target
  for the dual-channel architecture's explicit spin-coupling structure.
  (iv) \emph{Quantum-classical hybrid implementations.} Active spaces such
  as the iron--molybdenum cofactor of nitrogenase---a long-standing benchmark
  for fault-tolerant quantum algorithms~\cite{Reiher2017}---motivate workflows
  in which a classical Transformer sampler provides high-quality reference
  states that reduce the circuit depth required for quantum diagonalization.
  By preserving the handover mechanism of HI-VQE~\cite{PellowJarman2025,Yoo2026},
  HI-NQS retains a natural interface to future quantum-circuit samplers,
  where the classical Transformer can serve as either a noise-resilient
  baseline or a warm-start initialization for shorter-depth
  circuit-sampled SQD~\cite{RobledoMoreno2025,Yu2025}.

  More broadly, these benchmarks point to a structural shift: as active-space
  size grows, a classical generative sampler with inductive bias matched to
  fermionic spin structure can outpace perturbatively ordered SCI methods in
  basis compactness, suggesting that the gap between selected-CI and neural
  quantum states is a matter of architecture rather than principle.
  Recent NQS+SCI methods---NQS-SC~\cite{Solanki2026}, HAAR-SCI~\cite{HAAR2025},
  ARNN-SCI~\cite{Thompson2026}, GTNN-SCI~\cite{ShangGenCI2025}, and the
  QiankunNet family~\cite{Wu2023NNQS,ShangQKN2025,SuncuSCI2026}---are
  converging on a shared design principle: explicit feedback between a
  generative sampler and exact subspace diagonalization.
  Scaling-law analyses of Transformer-based NQS suggest that this class
  of ansatz has not yet reached its performance ceiling as a function of
  model size and active-space dimension~\cite{ScalingNQS2026}, reinforcing
  the case for continued architectural investment.
  HI-NQS contributes to this convergence by demonstrating that an inductive
  bias matched to the spin structure of fermionic Hamiltonians, when coupled
  to factorized eigenvector supervision, yields a sub-CIPSI scaling exponent
  on GPU hardware without any quantum-hardware resource.
  HI-NQS thus stands as a practical classical algorithm for strongly correlated
  quantum chemistry that preserves a direct interface to future hybrid
  classical--quantum workflows.

\section*{Competing Interests}
The authors declare no competing financial interest.

\section*{Data Availability}
The data supporting this study are available from the corresponding author upon reasonable request.

\section*{Acknowledgements}
J.-Y.~Chang, Y.-C.~Chang, Y.-J.~Lin, and M.-C.~Yang contributed equally to this work.
EJK acknowledges financial support from the National Science and Technology Council (NSTC) of Taiwan under Grant
No.~NSTC~114-2112-M-A49-036-MY3.

\bibliography{references}

\newpage
\noindent\rule{0.05in}{1.75in}%
\begin{minipage}[b][1.75in]{3.25in}
  \sffamily\frenchspacing
  \textbf{TOC Graphic placeholder.}\\
  Replace this box with \verb|\includegraphics{toc_graphic}|.\\
  JCTC recommended size: 8.5\,cm $\times$ 3.5\,cm.
\end{minipage}%
\rule{0.05in}{1.75in}

\end{document}